%% file: main-arXiv.tex
\documentclass[twoside,british,a4paper,twocolumn]{article}
\usepackage{amsmath}
\usepackage{amsfonts}
\usepackage{amssymb,upref}
\usepackage{tgheros} %% Main sans-serif font
\usepackage{tgtermes} %% Main serif font
\usepackage{anyfontsize}
\usepackage{xcolor}
\usepackage{graphicx}
\usepackage{enumitem}
\usepackage{subfig}
\usepackage{siunitx}

\usepackage{float}
\usepackage[hang,flushmargin]{footmisc}

\usepackage{balance}

\usepackage[utf8]{inputenc}
\usepackage[T1]{fontenc}

\usepackage{isodate}
\cleanlookdateon

\usepackage{lettrine}
\pdfmapfile{=montserrat.map} 

\usepackage{lipsum}

\usepackage{newtxmath}
\usepackage{geometry}
\usepackage{fancyhdr}

\fancypagestyle{firstpagestyle}
\usepackage{titlesec}
\titleformat{\section}[block]{\bfseries\upshape\sffamily\boldmath}{}{0.em}{}
\titlespacing*{\section}{0pt}{0.8em plus 0ex minus 0ex}{0em plus 0.ex}

\usepackage{titling}
\pretitle{
  \noindent
  \raggedright\huge\bfseries\upshape\sffamily\boldmath
  \color[rgb]{0,0,0.8}
  }
\posttitle{
  \vskip 0.6em
  }
\preauthor{
  \noindent 
  \large\mdseries\sffamily
  }
\postauthor{
  \vskip 0.4em
  {
  \sffamily
  \color{black!80}
  \noindent \small
  \address
  \par %\vskip -1.em
  \authoremail
  }
  \par \vskip 0.4em
  }
\predate{ \noindent \hspace{-0.4em} \sffamily\small}
\postdate{\vskip 0.0em}
\usepackage{caption}
\DeclareCaptionFont{cfs}{\fontsize{8.5}{10.25}\selectfont}
\DeclareCaptionLabelSeparator{vline}{\;|\;}
\DeclareMathAlphabet\mathbfcal{OMS}{cmsy}{b}{n}

\usepackage{tcolorbox}
\definecolor{abstractboxcolor}{cmyk}{0.1,0,0,0}
\newtcolorbox{abstractbox}{
  arc=0pt,
  boxrule=0pt,
  colback=abstractboxcolor,
  boxsep=0.5em,
  left=0pt, right=0pt, bottom=0pt, top=0pt,
  width=\columnwidth
}
\makeatletter
 \def\@textbottom{\vskip \z@ \@plus 1pt}
 \let\@texttop\relax
\makeatother
\usepackage[numbers,sort&compress]{natbib}

\makeatletter \def\NAT@def@citea{\def\@citea{\NAT@separator\,}} \makeatother % reduce spacing inside [1, 2].
\usepackage{etoolbox}
\apptocmd{\sloppy}{\hbadness 10000\relax}{}{}

\usepackage[%
  bookmarks=true,
  colorlinks,
  linkcolor=blue,
  urlcolor=blue,
  citecolor=blue,
  plainpages=false,
  pdfpagelabels,
  final,
  breaklinks=true
]{hyperref}
\hypersetup{
pdftitle={eFID}, 
pdfauthor={}
}

\usepackage{physics}

\renewcommand{\Re}{\operatorname{Re}}

\usepackage{textcomp}

\input{prephase}

\begin{document}

\twocolumn[
\begin{@twocolumnfalse}

\maketitle
\thispagestyle{firstpagestyle}

\vspace{-2mm}

\end{@twocolumnfalse}
]

\input{fid}

\balance

\bibliographystyle{arthur} 
\bibliography{lit}

\hfill 

\clearpage

\newpage

\input{methods}

\newpage

\onecolumn

{
\raggedright\huge\bfseries\upshape\sffamily\boldmath
\color[rgb]{0,0,0.8}
Supplementary Information
}

\input{supplementary}

\end{document}

%% file: prephase.tex
\title{Enantiosensitive steering of free-induction decay}
\newcommand\shorttitle{Enantiosensitive steering of free-induction decay}

\author{Margarita Khokhlova\textsuperscript{1$\,*$}, Emilio Pisanty\textsuperscript{1}, Serguei Patchkovskii\textsuperscript{1}, Olga Smirnova\textsuperscript{1,2}, and Misha Ivanov\textsuperscript{1,3,4}}
\newcommand\shortauthor{M. Khokhlova, E. Pisanty, S. Patchkovskii, O. Smirnova, and M. Ivanov}
\newcommand\address{\textsuperscript{1}Max Born Institute, Berlin 12489, Germany \\
\textsuperscript{2}Technische Universität Berlin, Berlin, Germany \\
\textsuperscript{3}Department of Physics, Humboldt University, Berlin, Germany \\
\textsuperscript{4}Blackett Laboratory, Imperial College London, London, UK \\}
\newcommand\authoremail{$^*$m.khokhlova@imperial.ac.uk}

\date{ \today }

%% file: fid.tex
%%% Abstract %%%
%
{\bf \noindent
Chiral discrimination, a problem of vital importance~\cite{Palyi2019}, has recently become an emerging frontier in ultrafast physics~\cite{Lux2012, Beaulieu2018, Cireasa2015, Ayuso2019, Neufeld2021}, with remarkable progress achieved in multiphoton~\cite{Lux2012,Beaulieu2018} and strong-field~\cite{Cireasa2015, Ayuso2019, Neufeld2021, Neufeld2019, Ayuso2020} regimes.
Rydberg excitations, unavoidable in the strong-field regime and intentional for few-photon processes~\cite{Lux2012, Beaulieu2018}, arise in all these approaches. 
Here we show how to harness this ubiquitous feature by introducing a new phenomenon, enantiosensitive free-induction decay, steered by a tricolour chiral field at a gentle intensity, structured in space and time. 
We demonstrate theoretically that an excited chiral molecule accumulates an enantiosensitive phase due to perturbative interactions with the tricolour chiral field, resulting in a spatial phase gradient steering the free-induction decay in opposite directions for opposite enantiomers.
Our work introduces a general, extremely sensitive, all-optical, enantiosensitive detection technique which avoids strong fields and takes full advantage of recent advances in structuring~light.
}
%

%%% Background %%%

Chiral recognition is an essential task in chemistry, whose origin dates back to the birth of the discipline~\cite{Pasteur1905} with the discovery of the optical activity of biomolecules: in solution, different enantiomers, which are non-superimposable mirror images of each other, can rotate in opposite directions the polarisation of light that travels through the medium. 
However, for dilute media and in gas phase, this effect is severely challenging to implement, since it relies on rather weak optical magnetic interactions.
This creates a strong demand for an optical chiral discrimination method which relies purely on dipole-interaction physics and is based only on the local electric field of the light~--- a goal which is not possible to achieve within linear optics~\cite{Tang2010}.

A recent breakthrough in nonlinear optics has bypassed this barrier, showing that nonlinear optical processes can work as a key to molecular chirality using only electric-field optical interactions~\cite{Ayuso2019, Neufeld2019, Ayuso2020, Ordonez2018}. This revolution improves on previous work on chiral nonlinear optics~\cite{Rentzepis1966, Fischer2000, Simpson2004, Belkin2005, Patterson2013, Chen2020, Eibenberger2017} and has been accompanied by a wealth of other electric-field methods coupled to other non-optical observables~\cite{Lux2012, Beaulieu2018, Ordonez2019ii, Cireasa2015, Pitzer2013, Perez2017, Neufeld2021}.
These approaches are generally constructed using the tools of strong-field and ultrafast physics, and they can therefore naturally take advantage of the rich toolbox of attosecond science~\cite{Krausz2009, Villeneuve2018, Biegert2021}.
However, this advantage has come at the price of nonperturbative processes that require high intensities~\cite{Ayuso2019, Neufeld2019, Ayuso2020}, which creates the need for delicate schemes that apply the recently-discovered nonlinear chiral properties of the spatially-structured electromagnetic field~\cite{Ayuso2019, Neufeld2021} without destroying or disturbing the molecule.

%%% Our intro %%%

\begin{figure*}[h]
    \centering
    \includegraphics[width=0.90\linewidth]{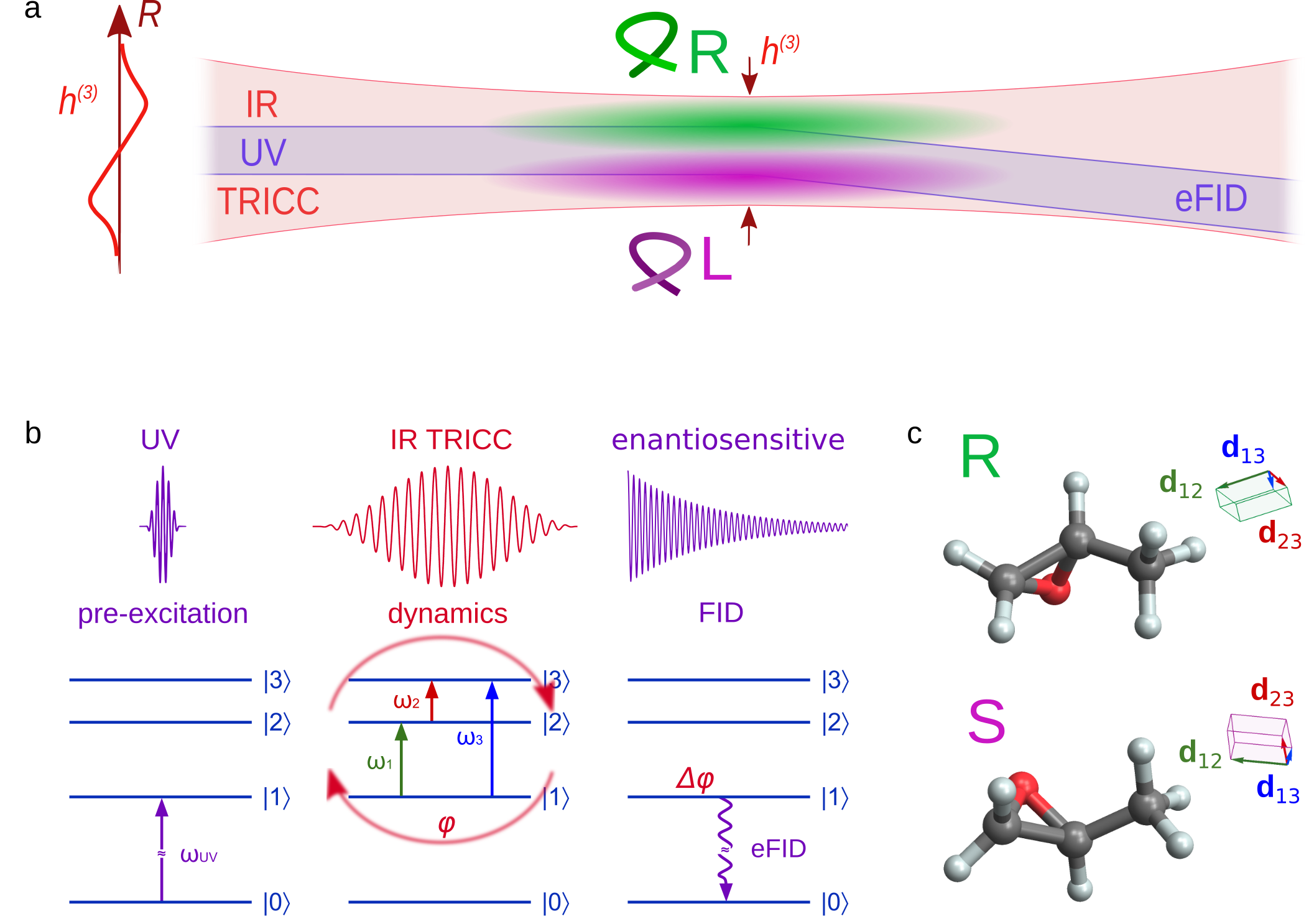}
    \caption{Enantiosensitive FID. 
    (a) Focused UV and tricolour IR beams interact with molecules in the focal region. Due to the focusing, this field becomes chiral, with the chiral correlation function $h^{(3)}$ changing its sign across the focus (green and purple shades). The profile of $h^{(3)}$ is shown on the left; the slope of $h^{(3)}$ at the optical axis results in the redirection of the FID beam by chiral molecules. 
    (b) Scheme of the single-particle interaction. On the first step, the UV pre-excites the molecule from the ground state $|0 \rangle$ into the excited state $|1 \rangle$. The TRICC field then induces dynamics between the excited states $|1 \rangle$, $|2 \rangle$ and $|3 \rangle$, resulting in an enantiosensitive quantum phase $\Delta \varphi$ of the FID-active state $|1 \rangle$ after the TRICC pulse. 
    (c) R-methyloxirane and S-methyloxirane, and their three main transition dipoles, forming opposite chiral triplets.
    }
    \label{fig:eFIDscheme}
\end{figure*}

In this Letter we propose an experiment allowing chiral recognition on an ultrafast timescale using non-destructive weak fields. Our scheme builds on the recent advances on chiral synthetic light~\cite{Ayuso2019} to induce a controllable enantiosensitive quantum phase of the medium, which is then translated into easily measurable macroscopic observables by leveraging the exquisite control over light afforded by present progress in structured light~\cite{Rubinsztein2016} and in ultrafast control~\cite{Krausz2009, Villeneuve2018, Biegert2021}. 
Specifically, we adapt the ability to steer free-induction decay (FID) radiation, recently demonstrated via quantum phase manipulation in atomic gases~\cite{Bengtsson2017}, to chirally-sensitive drivers interacting with chiral media (see Figure~\ref{fig:eFIDscheme}a), thereby introducing an enantiosensitive Stark shift which gives rise to FID labelling of enantiomers (FIDLE).\!%
\footnote{%
Note that our approach, based on pure electric dipole physics, is distinct from previous observations of chiral FID in dense media based on magnetic interactions~\cite{Ghosh2021}.
}

At the microscopic level, we model a chiral molecule promoted from its ground state $|0 \rangle$ into an FID-active excited vibronic state $|1 \rangle$ by a coherent pump~--- in our case, a short UV pulse (see Figure~\ref{fig:eFIDscheme}b)~--- and which re-emits this photon energy by decaying back to the ground state.
The phase of this emission, which is responsible for the direction of the resulting FID beam, is defined by the quantum phase of the FID-active state.
The enantiosensitive contribution to the quantum phase is imparted by the enantiosensitive Stark shift arising in synthetic chiral fields. The simplest way to introduce the enantiosensitive Stark shift is to consider non-resonant interaction of the FID-active state $|1 \rangle$ with two other excited states, $|2 \rangle$ and $|3 \rangle$, induced by a tricolour chiral (TRICC) combination of IR fields with frequencies $\omega_1$, $\omega_2$ and $\omega_3=\omega_1+\omega_2$ and noncollinear polarisations forming a chiral triplet.
We create this chiral triplet macroscopically using tightly-focused Gaussian beams to provide a longitudinal polarisation component~\cite{Bliokh2015}, resulting in a chiral time evolution of the electric field at every point.
This construction corresponds to a field chirality which changes sign across the focus, and this sign change is directly converted into the quantum phase of the FID-active state, thus steering the FID emission.

To demonstrate this, we first develop a general analytical theory of the chiral sensitivity of the TRICC FIDLE dynamics, and we suggest field configurations that produce the locally-chiral electric fields required to drive these dynamics.
We then present simulations of both near- and far-field observables for the methyloxirane molecule, showing a clearly visible enantiosensitive steering of the FID emission. Finally, we explore the effect of the various TRICC-field parameters on this steering.
Additional details and benchmarking are included in the Supplementary Information (SI).

%%% Theory bit %%%

The core of our scheme is the chirally-sensitive dynamics driven by the TRICC-field combination. 
We introduce these dynamics using a simple model of a molecule with three states involved, as shown Figure~\ref{fig:eFIDscheme}b.
The system is driven by the TRICC pulse, consisting of three fields
\begin{equation}
\mathbfcal{E}= \Re\mathopen{}\left[\sum\limits_{j=1}^{3}\mathbfcal{E}_j e^{-i(\omega_j t + \phi_j)} \right]\mathclose{}\, 
\label{tricc_field} 
\end{equation}
(described in detail below),
which induces transitions between each couple of states $|1\rangle$, $|2\rangle$ and $|3\rangle$.
The time-dependent Schr\"odinger equation (TDSE) is solved fully analytically in both the resonant and off-resonant cases (see Methods and SI). In both cases the complex amplitude of the FID-active state is presented in the form $c_1(t)=e^{i \delta E t}$, in the interaction picture with respect to the molecular Hamiltonian, where $\delta E$ is an energy shift resulting in the phase $\Delta\varphi=\int_0^{\tau} \delta E dt$ accumulated during the TRICC pulse with duration $\tau$. 

In the off-resonant case the energy shift is written as 
\begin{equation}
\delta E = 2 \frac{|V_{12}| |V_{13}| |V_{23}|}{\omega_{12} \omega_{13}} \cos{\phi} +\frac{|V_{12}|^2}{\omega_{12}} + \frac{|V_{13}|^2}{\omega_{13}}  \, ,
\label{off_res_en_shift}
\end{equation}
where $V_{12} = \mathbf{d}_{12} \cdot \mathbfcal{E}_1 e^{i\phi_1}/2$, $V_{23} = \mathbf{d}_{23} \cdot \mathbfcal{E}_2 e^{i\phi_2}/2$ and $V_{13} = \mathbf{d}_{13} \cdot \mathbfcal{E}_3 e^{i\phi_3}/2$ are dipole-interaction matrix elements, $\omega_{12}=E_2-E_1-\omega_1$, $\omega_{23}=E_3-E_2-\omega_2$ and $\omega_{13}=E_3-E_1-\omega_3$ are detunings from exact resonances, and $\phi=\phi_1 + \phi_2 - \phi_3$ is the relative TRICC phase. 
After molecular orientation averaging (see Methods) the energy shift becomes
\begin{equation}
\begin{split}
\langle \delta E \rangle_{\mathcal{O}} =& \frac{\Re  \mathopen{} \left\{ \left( \mathbf{d}^*_{13} \cdot \left[ \mathbf{d}_{12} \times \mathbf{d}_{23} \right] \right) \left( \mathbfcal{E}^*_3 \cdot \left[\mathbfcal{E}_1 \times \mathbfcal{E}_2 \right] e^{i\phi} \right)\mathclose{} \right\}}{24 \omega_{12} \omega_{13}}  \\ 
& \qquad \qquad \qquad+\frac{|\mathbf{d}_{12}|^2 |\mathbfcal{E}_1|^2}{12\omega_{12}} + \frac{ |\mathbf{d}_{13}|^2 |\mathbfcal{E}_3|^2}{12 \omega_{13}}  \, .
\end{split}
\label{off-res_en_shift_aver}
\end{equation}
The two last terms represent the ordinary Stark shift of the FID-active state~\cite{Bengtsson2017}, whereas the first term, the enantiosensitive Stark shift, has an explicit separation into a triple product of molecular dipoles, which carries information about the chirality of the molecule, and a triple product of the driving fields. 
This term is non-zero if the molecule is chiral and it is driven by an electrically-chiral field, and therefore it provides the enantiosensitivity of the FID-beam steering of our scheme.

\begin{figure*}[ht!]
    \centering
    \includegraphics[width=0.98\linewidth]{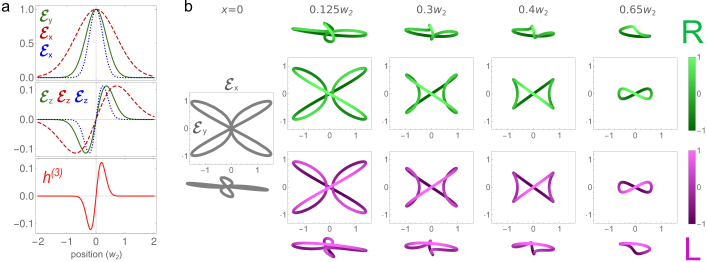}
    \caption{TRICC field. 
    (a) Field components of each $\omega_1$, $\omega_2$ and $\omega_3$ TRICC-field colour (solid green, dashed red and dotted blue, respectively) both transverse as a function of $x$ (top) and longitudinal as a function of $x$ for $\omega_2$, $\omega_3$ and $y$ for $\omega_1$ (middle), producing a nonzero chiral correlation function $h^{(3)}$ (bottom).
    The fields are normalised to the central value of the transverse component.
    (b) 3D Lissajous figures of the TRICC field, forming a `chiral clover', at different positions along the $x$ axis. The first column (gray) shows the achiral case $x=0$. For $x \neq 0$, the top two rows (green) correspond to positive values of $x$ and the bottom two rows (lilac) to negative values. The middle two rows are the projections of the Lissajous figures on the $xy$-plane, and bottom and top show an angled viewpoint. 
    The lightness of the curve (colour scales on the right) represents the value of the longitudinal component $\mathcal{E}_z(t)$.
    We show fields in a $\omega_1{:}\omega_2{:}\omega_3 = 2{:}1{:}3$ configuration with wavelengths $\lambda_1=\SI{800}{nm}$, $\lambda_2=\SI{1600}{nm}$ and $\lambda_3=\SI{533}{nm}$ and phases $\phi_1=\pi/3$, $\phi_2=-\pi/3$ and $\phi_3=\pi$, focused to $w_i=1.2\lambda_i$ with equal numerical aperture for all three colours.
    An alternative set of Lissajous figures, showing knotted polarisations~\cite{Sugic2020}, is included in the SI.
   }
    \label{fig:tricc}
\end{figure*}

%%% Field description %%%

\begin{figure}[h!]
    \begin{tabular}{c}
    $\ $ \\
    $\ $ \\
    \includegraphics[width=0.9\linewidth]{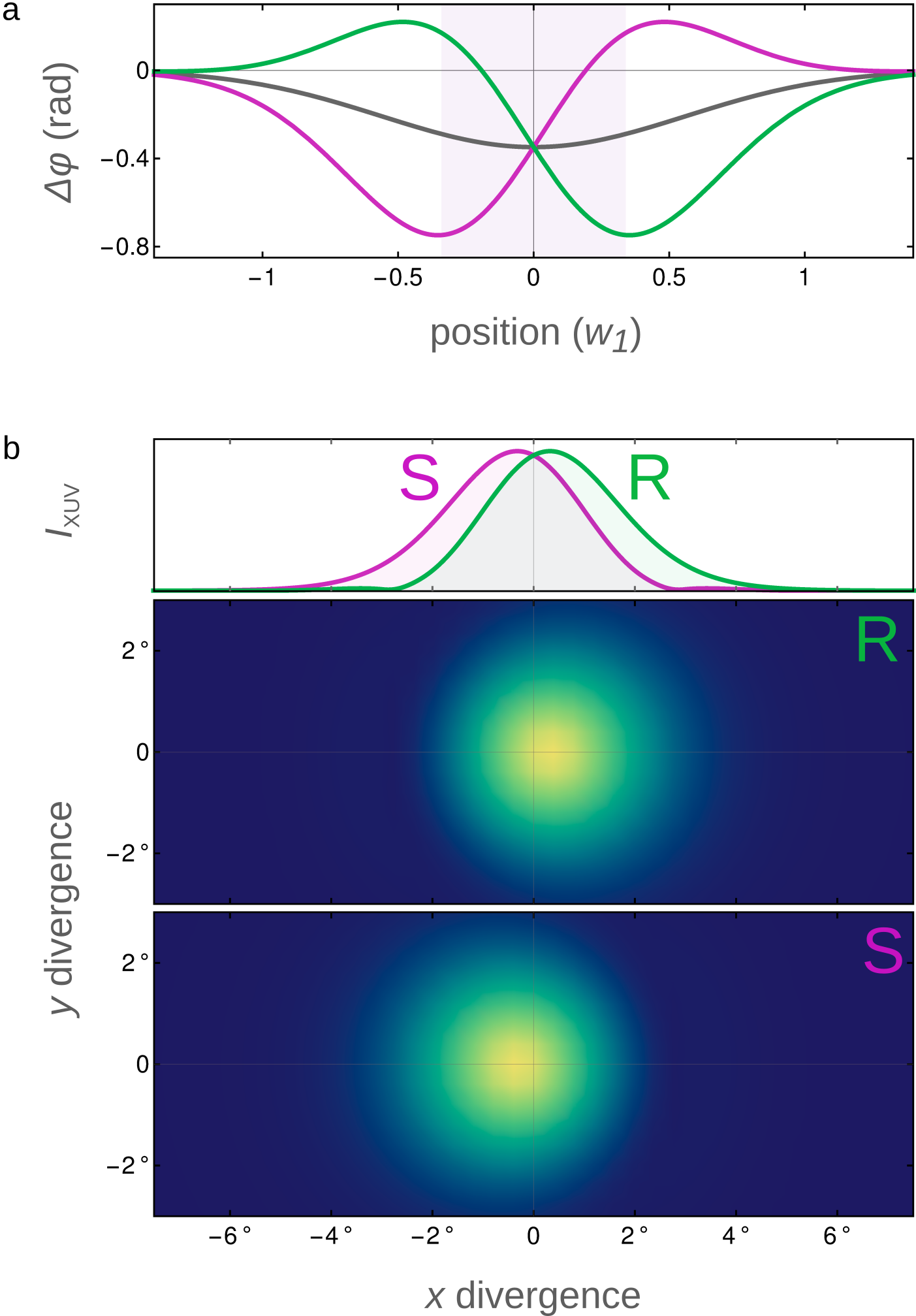}
    \\ $\ $
    \end{tabular}
    \caption{FIDLE by methyloxirane. 
    (a) Phase $\Delta \varphi$ accumulated in the FID-active Rydberg $3s$ state for R (green) and  S (lilac) enantiomers; gray shows an achiral phase. 
    (b) FID beam divergence for each enantiomer, with a lineout of both on the top panel. 
    }
    \label{fig:real_res}
\end{figure}

The field triple-product represents the leading-order nonlinear chiral correlation function of the field~\cite{Ayuso2019},
\begin{equation}
h^{(3)}
=
\mathbfcal{E}_3^* \cdot \left( \mathbfcal{E}_1 \times \mathbfcal{E}_2 \right) e^{i\phi} \, ,%.
\label{h3-definition}
\end{equation}
%This is the core component of the Fourier transform of the time-domain three-point correlation function $H^{(3)}(\tau_1,\tau_2) = \int \mathbfcal{E}(t) \cdot \left( \mathbfcal{E}(t+\tau_1) \times \mathbfcal{E}(t+\tau_2) \right) \mathrm{d}t$, and it 
which provides a quantitative measure of the chirality of the Lissajous curve traced out by the TRICC electric field over time.
Such a quantitative measure cannot be provided within traditional understandings of optical chirality~\cite{Tang2010}, since linear optics is blind to sub-cycle dynamics. In our case, the correct order of nonlinearity is the one dictated by the physics of the TRICC process, a fact reflected in the natural appearance of $h^{(3)}$ in the orientation-averaged energy shift~\eqref{off-res_en_shift_aver}.

\begin{figure*}[ht!]
    \centering
    \vspace{3mm}
    \includegraphics[width=0.98\linewidth]{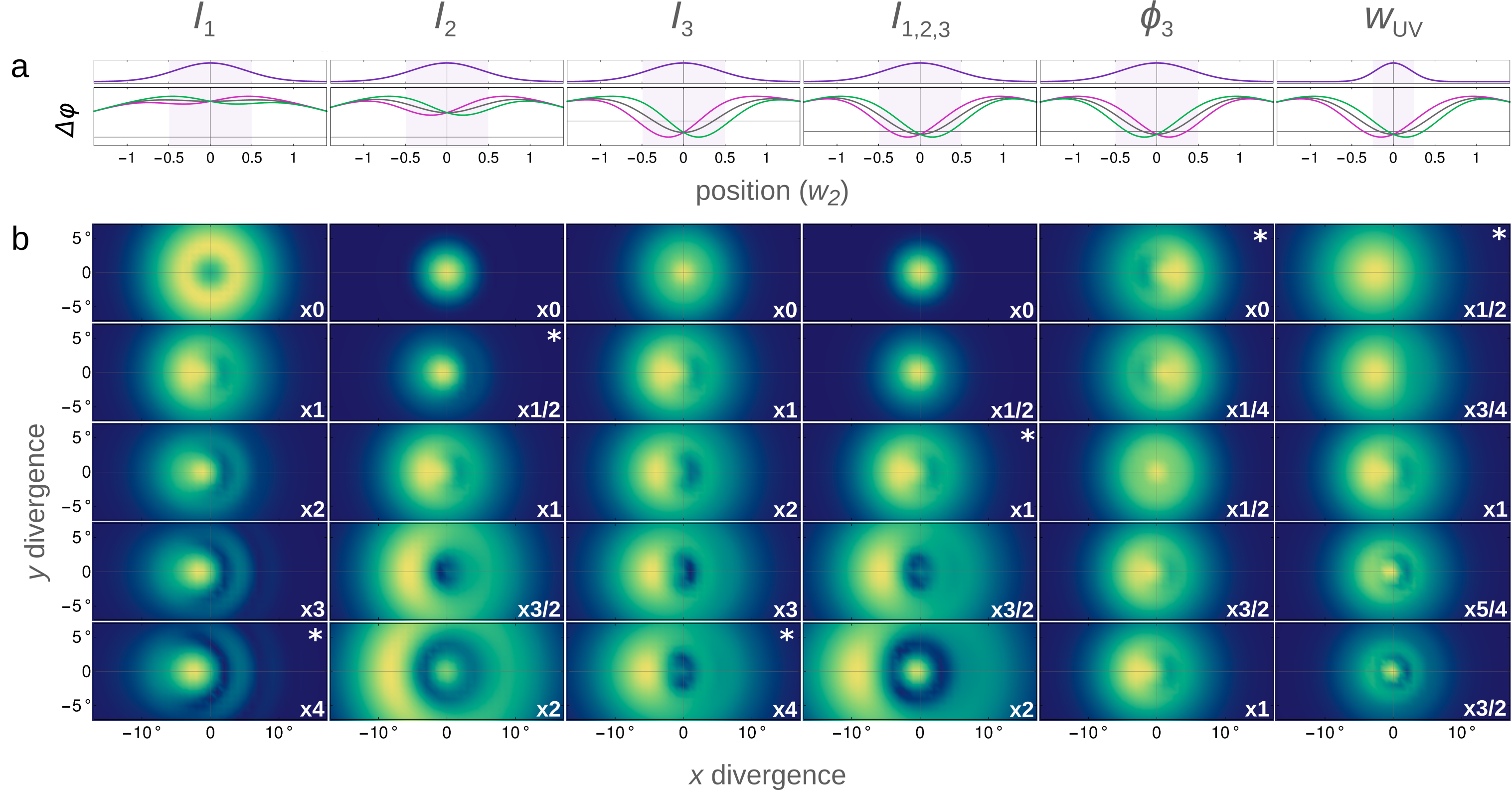}
    \caption{
    Enantiosensitive FID steering as a function of the TRICC field.
    Parameter scans reflecting the change in (a) the accumulated phase $\Delta \varphi$ of the FID-active state, and (b) the final divergence of the FID beam, as displayed in Figure~\ref{fig:real_res}, for the S enantiomer. 
    The scanned parameter is shown above each column, with asterisks showing the panel corresponding to the phase plot in (a). 
    The overall plot parameters use intensities $I_1= \SI{0.5e10}{W/cm^2}$, $I_2= \SI{8e11}{W/cm^2}$ and $I_3= \SI{1e11}{W/cm^2}$, phase $\phi_3=\pi$ and UV focal waist $w_\mathrm{UV} = 4 \lambda_\mathrm{UV}$. 
    For each column, the scanned parameter is changed via the multiplier at bottom right of each panel.
   }
    \label{fig:carpet}
\end{figure*}

The TRICC field~\eqref{tricc_field} itself is a superposition of three components 
%of the form
% \begin{equation}
% \mathbfcal{E}= \Re\mathopen{}\left[\sum\limits_{j=1}^{3}\mathbfcal{E}_j e^{-i(\omega_j t + \phi_j)} \right]\mathclose{}\, ,
% \label{tricc_field} 
% \end{equation}
with amplitudes $\mathbfcal{E}_j$ and frequencies $\omega_j$.
%In the far field~--- that is, 
Before the focusing optics that directs the TRICC beams onto the interaction region~--- the beams are polarised in the plane perpendicular to the propagation direction, with two of these components orthogonal to the third one: $\mathbfcal{E}_1=\{0,\mathcal{E}_1,0 \}$, $\mathbfcal{E}_2=\{\mathcal{E}_2,0,0 \}$ and $\mathbfcal{E}_3=\{\mathcal{E}_3,0,0 \}$. In the near field each of these components forms a Gaussian focus, which acquires a longitudinal polarisation component within the first post-paraxial approximation~\cite{Bliokh2015}, giving field amplitudes with spatial dependence of the form
\begin{equation}
\begin{split}
\mathbfcal{E}_1 = & i \sqrt{I_1} e^{-(x^2+y^2)/w_1^2} \left\{0,1,-i\tfrac{2y}{k_1 w_1^2} \right\} \, , \\
\mathbfcal{E}_2 = & i \sqrt{I_2} e^{-(x^2+y^2)/w_2^2} \left\{1,0,-i\tfrac{2x}{k_2 w_2^2} \right\} \, , \\
\mathbfcal{E}_3 = & i \sqrt{I_3} e^{-(x^2+y^2)/w_3^2} \left\{1,0,-i\tfrac{2x}{k_3 w_3^2} \right\} \, 
\end{split}
\label{tricc_amps}
\end{equation}
at the focal plane $z=0$, where $I_j$ are field intensities at the centre of the focal spot, $k_j=\omega_j/c$ are wavenumbers, and $w_j$ are focal waists.

Figure~\ref{fig:tricc}a describes the main features of the TRICC field at the focal plane, with the transverse polarisation components (upper panel) inducing a longitudinal component (middle panel), in post-paraxial optics, to comply with the Maxwell equations in free space, $\nabla\cdot\mathbfcal{E}=0$.
This longitudinal component provides the three-dimensionality necessary to produce a nonzero chiral correlation function $h^{(3)}$, shown in the lower panel.

In the time domain, the three-dimensional polychromatic field combination~\eqref{tricc_field} traces a chiral Lissajous figure over time, shown in Figure~\ref{fig:tricc}b, with opposite chiralities on either side of the optical axis, shown in green ($x>0$) and lilac ($x<0$) respectively.
These chiral Lissajous figures form enantiomeric pairs: they are exact mirror images of each other, but they cannot be superimposed on each other using only rotations, and, as shown in the lower panel of Figure~\ref{fig:tricc}a, they correspond to opposite signs of the chiral correlation function $h^{(3)}$.
At the centre of the beam ($x=0$), the longitudinal polarisation components vanish, leading to a planar Lissajous figure which is therefore achiral.

%%% Main result %%%

To illustrate the mechanics of TRICC enantiosentitive steering more vividly, we now turn to a realistic example. We use as a benchmark the methyloxirane molecule,\!%
\footnote{%
Also denoted as propylene oxide and epoxypropane in strong-field literature; we use the preferred IUPAC designation.
} %
shown in Figure~\ref{fig:eFIDscheme}c, which is a common choice due to its small size and relative rigidity~\cite{Barone2014}.
For a realistic molecule, we extend the three-level model system to account for all the relevant transitions, producing a generalisation of~\eqref{off-res_en_shift_aver} discussed in the SI.
All of the molecular parameters, including eigenstate energies and transition dipole matrix elements are obtained from \emph{ab initio} calculations (see Methods) and are detailed in the SI.

We show in Figure~\ref{fig:real_res} the enantiosensitive beam steering that results, for both near- and far-field observables. For the near-field, Figure~\ref{fig:real_res}a displays the quantum phase $\Delta\varphi$ induced in the FID-active state of the molecule, for both the S (lilac curve) and R (green curve) enantiomers, as well as for the achiral state (gray curve) obtained from~\eqref{off-res_en_shift_aver} by removing the chirally-sensitive terms. This quantum phase experiences a clear slope over the focal region of the UV beam (shaded in violet) which directly mirrors the chiral nonlinear correlation function $h^{(3)}$ of the field, and which redirects the beam in opposite directions for different chiralities.%
\!\footnote{%
For a racemic mixture, the opposite phases cancel out, and the beam is not deflected. For an impure but imbalanced mixture, the phase is averaged, providing the basis to use this scheme for measurements of enantiomeric purity.
}
This redirection translates into the far-field picture, which we exhibit in Figure~\ref{fig:real_res}b both as a line-out and a 2D image, via a standard spatial Fourier transform. The two beams perform a clearly visible displacement at about half a degree from the initial central position.

This result is obtained for a TRICC field at wavelengths $\lambda_1 = \SI{3438}{nm}$, $\lambda_2 = \SI{1365}{nm}$ and $\lambda_3 = \SI{977}{nm}$, chosen to be close to resonances with the Rydberg $3p$ and $3d$ states (see SI for details and for an alternative choice of wavelengths).
We use field intensities $I_1=\SI{2e10}{W/cm^2}$, $I_2= \SI{3.5e11}{W/cm^2}$ and $I_3= \SI{2e11}{W/cm^2}$, phases as in Figure~\ref{fig:tricc}, and a $\sin^2$ envelope starting at $t=0$ lasting 25 optical cycles (FWHM of amplitude) of the $\omega_2$ field.
The UV beam is focused to a waist of $w_\mathrm{UV}=7 \lambda_\mathrm{UV}$ (FWHM) and the TRICC fields are focused to equal waists $w_1=w_2=w_3=1.2\lambda_1$. 
These TRICC-field parameters are chosen in such a way to minimise population transfer to higher excited states and to keep the TRICC pulse short enough to avoid triggering nuclear dynamics. We benchmark these results against a direct numerical solution of the TDSE reported in the SI.

%%% Carpet %%%

To understand more widely how the FIDLE beam steering works, it is also useful to look at broader variations in these parameters. We show this in Figure~\ref{fig:carpet} for a three-state model of methyloxirane taking only the states closest to resonances with fields of wavelengths $\lambda_1 = \SI{3.4}{\mu m}$, $\lambda_2 = \SI{1.35}{\mu m}$, and $\lambda_3 = \SI{966}{nm}$, otherwise using the same parameters as in Figure~\ref{fig:real_res}. The `carpet' in Figure~\ref{fig:carpet} shows the effect of variations in the driver intensities as well as the relative TRICC-field phase and the UV focal waist.

This `carpet' demonstrates that the enantiosensitive FID steering is possible at various intensities and with varying degrees of induced structure in the FID beam. 
We see, in particular, that changing the TRICC-field intensities directly controls the magnitude of the beam divergence angle, whereas the relative phase (given by the $\phi_3$ scan) also controls the direction of the steering. This is natural, since the phase appears in the chiral correlation function $h^{(3)}$ defined in~\eqref{h3-definition}, and changing $\phi$ by $\pi$ inverts the chirality of the field and therefore it inverts its interaction with chiral matter, so the FID beam is redirected in the opposite direction for each fixed enantiomer.

%%% Conclusion %%%

The process we propose provides for clear enantiosensitive signals from a dilute medium, by harnessing the power of synthetic chiral light while still keeping to a delicate intensity, thereby preserving the molecules largely undisturbed. Our scheme is compatible with a wide array of ultrafast pump-probe spectroscopies and it is built on sources and techniques which are already available. This thus opens the door to widespread application of chiral spectroscopies in ultrafast science, and it supplies a general tool for optical chiral recognition.

%% file: methods.tex
\section{Methods}
\subsubsection*{TRICC-driven dynamics of molecular states}
We solve the TDSE, using atomic units throughout,
\begin{equation}
i\frac{\partial \Psi(t)}{\partial t}=\hat{H}(t) \Psi(t)
\label{TDSE}
\end{equation}
for the Hamiltonian $\hat{H}(t)=\hat{H}_0+\hat{V}(t)$, where $\hat{H}_0$ is the unperturbed Hamiltonian of the molecule $\hat{H}_0\varphi_n=\tilde{E}_n \varphi_n$, $n=1,2,3$, and $E_n$ are the energies of the excited states, assuming that the molecule is in the FID-active state $|1 \rangle$ at the start of the TRICC pulse. The term $\hat{V}(t)=- \mathbf{d} \cdot \mathbfcal{E}$ describes the interaction the TRICC laser field~\eqref{tricc_field} with the electric field amplitudes $\mathbfcal{E}_j$, frequencies $\omega_j$, and phases $\phi_j$. 

The total wave function $\Psi(t)$ is given by
\begin{equation}
\Psi(t)=\sum\limits_{n=1}^3 c_n(t)\varphi_n e^{-iE_n t} \, ,
\label{tot_wf}
\end{equation}
where $c_n$ are complex amplitudes of the states $|1 \rangle$, $|2 \rangle$ and $|3 \rangle$ with energies $E_n$, respectively, (see Fig.~\ref{fig:eFIDscheme}b).

We substitute the total wave function~(\ref{tot_wf}) into the TDSE~(\ref{TDSE}), and by using the standard rotating-wave approximation obtain the usual system of differential equations for the complex amplitudes
\begin{equation}
\begin{split}
i \dot{c}_1=& - V_{12} e^{-i\omega_{12} t} c_2 - V_{13} e^{-i\omega_{13} t} c_3 \, ,\\
i \dot{c}_2=& - V^*_{12} e^{i\omega_{12} t} c_1 - V_{23} e^{-i\omega_{23} t} c_3 \, ,\\
i \dot{c}_3=& - V^*_{13} e^{i\omega_{13} t} c_1 - V^*_{23} e^{i\omega_{23} t} c_2 \, ,\\
\end{split}
\label{general_system}
\end{equation}
where 
\begin{equation}
\begin{split}
V_{12} = \mathbf{d}_{12} \cdot \mathbfcal{E}_1 e^{i\phi_1}/2 \, , \\ 
V_{23} = \mathbf{d}_{23} \cdot \mathbfcal{E}_2 e^{i\phi_2}/2 \, , \\ 
V_{13} = \mathbf{d}_{13} \cdot \mathbfcal{E}_3 e^{i\phi_3}/2 \,
\end{split}
\label{mat_els}
\end{equation}
are interaction matrix elements including the dipole transition matrix elements $d_{ij}=\langle \varphi_i|\mathbf{d}|\varphi_j\rangle$, and
\begin{equation}
\begin{split}
\omega_{12}=E_2-E_1-\omega_1 \, , \\ 
\omega_{23}=E_3-E_2-\omega_2 \, , \\ 
\omega_{13}=E_3-E_1-\omega_3 \, ,
\end{split}
\label{detunings}
\end{equation}
are detunings from the exact resonance.

The solution for the case of the exact resonance of this system is detailed in the corresponding section in SI. 

In the off-resonant case this system can be solved within perturbation theory (PT) up to the third order assuming that the main population is concentrated at the lowest excited level $|1 \rangle$. The first step of PT gives
\begin{equation}
\begin{split}
i \dot{c}_2=& - V^*_{12} e^{i\omega_{12} t} c_1 \, ,\\
i \dot{c}_3=& - V^*_{13} e^{i\omega_{13} t} c_1 \, ,\\
\end{split}
\label{pt1}
\end{equation}
or
\begin{equation}
\begin{split}
c_2=& - \frac{V^*_{12}}{\omega_{12}} e^{i\omega_{12} t} c_1 \, ,\\
c_3=& - \frac{V^*_{13}}{\omega_{13}} e^{i\omega_{13} t} c_1 \, .\\
\end{split}
\label{pt1_integr}
\end{equation}
At the second step of PT, we have
\begin{equation}
\begin{split}
i \dot{c}_2=& - V^*_{12} e^{i\omega_{12} t} c_1 -\frac{V^*_{13}V_{23}}{\omega_{13}} e^{i(\omega_{13}-\omega_{23}) t} c_1 \, ,\\
i \dot{c}_3=& - V^*_{13} e^{i\omega_{13} t} c_1 -\frac{V^*_{12}V^*_{23}}{\omega_{12}} e^{i(\omega_{12}+\omega_{23}) t} c_1 \, ,\\
\end{split}
\label{pt2}
\end{equation}
or
\begin{equation}
\begin{split}
c_2=& \left[ \frac{V^*_{12}}{\omega_{12}} + \frac{V^*_{13}V_{23}}{\omega_{12}\omega_{13}} \right] e^{i\omega_{12} t} c_1 \, ,\\
c_3=& \left[ \frac{V^*_{13}}{\omega_{13}} + \frac{V^*_{12}V^*_{23}}{\omega_{12}\omega_{13}} \right] e^{i\omega_{13} t} c_1 \, .\\
\end{split}
\label{pt2_integr}
\end{equation}
Finally, the third step brings us to the dynamics of the lowest excited state:
\begin{equation}
c_1 (t)=  e^{
i 
\left[
\frac{|V_{12}|^2}{\omega_{12}} + \frac{|V_{13}|^2}{\omega_{13}} 
 + 2 \frac{|V_{12}||V_{23}||V_{13}|}{\omega_{12}\omega_{13}} \cos{\phi} 
 \right] 
 t 
 }  \, , 
\label{pt3_integr}
\end{equation}
where $\phi$ is a relative TRICC phase.

%\newpage

\subsubsection*{Orientation averaging}
The full optical response is the average of the phase of the emission over all possible molecular orientations, 
a critical step in comparing the optical response of opposite enantiomers (which is missing in previous related work~\cite{Chen2020}).
For the analytical result in \eqref{pt3_integr}, this can be calculated exactly using the theory of isotropic tensors.

For the Stark shifts $\frac{|V_{12}|^2}{\omega_{12}}$ and $\frac{|V_{13}|^2}{\omega_{13}}$, the orientation-averaged amplitudes simplify to
\begin{align}
\langle|V_{12}|^2\rangle_\mathcal{O}
& =
\langle (\mathbf{d}_{12} \cdot \mathbfcal{E}_1 e^{i\phi_1}/2) (\mathbf{d}_{12} \cdot \mathbfcal{E}_1 e^{i\phi_1}/2)^* \rangle_\mathcal{O}
\\ & =
\frac{1}{4}
\mathcal{E}_{1,i}^{\phantom{*}} \mathcal{E}_{1,j}^*
\langle {d}_{12,i}^{\phantom{*}} {d}_{12,j}^* \rangle_\mathcal{O}
\end{align}
using Einstein summations. Here the orientation average of the molecular tensor ${d}_{12,i}^{\phantom{*}} {d}_{12,j}^*$ reduces it to an isotropic tensor, which must be of the form $\langle {d}_{12,i}^{\phantom{*}} {d}_{12,j}^* \rangle_\mathcal{O} = D_2 \, \delta_{ij}$, where the multiplier $D_2$ is determined by taking the trace:
\begin{equation}
3D_2
= D_2 \, \delta_{ii}
= \langle {d}_{12,i}^{\phantom{*}} {d}_{12,i}^* \rangle_\mathcal{O}
= |\mathbf{d}_{12}|^2,
\end{equation}
and the isotropic $\delta_{ij}$ produces the inner product $\delta_{ij} \, \mathcal{E}_{1,i}^{\phantom{*}} \mathcal{E}_{1,j}^*  = |\mathbfcal{E}_1|^2$, giving the final result
\begin{equation}
\langle|V_{12}|^2\rangle_\mathcal{O}
=
\frac{1}{12}
|\mathbf{d}_{12}|^2
|\mathbfcal{E}_1|^2.
\end{equation}

For the chirally-sensitive triple product, we write 
\begin{equation}
|V_{12}||V_{23}||V_{13}|\cos{\phi}
=
\frac{1}{8}
\Re\mathopen{}\left[
(\mathbf{d}_{12} {\cdot} \mathbfcal{E}_1)
(\mathbf{d}_{23} {\cdot} \mathbfcal{E}_2)
(\mathbf{d}_{13}^* {\cdot} \mathbfcal{E}_3^*)
e^{i\phi}
\right]\mathclose{}
\end{equation}
so, similarly to the Stark shifts, we can separate
\begin{align}
\left\langle
|V_{12}||V_{23}||V_{13}|\cos{\phi}
\right\rangle_\mathcal{O}
& =
\frac{1}{8}
\Re\mathopen{}\left[
\mathcal{E}_{1,i}^{\phantom{*}} \mathcal{E}_{2,j}^{\phantom{*}} \mathcal{E}_{3,k}^* e^{i\phi}
\nonumber \right. \\ & \qquad \qquad  \left.
\left\langle
{d}_{12,i}^{\phantom{*}} {d}_{23,j}^{\phantom{*}} {d}_{13,k}^* 
\right\rangle_\mathcal{O}
\right]\mathclose{}.
\end{align}
Here the molecular tensor ${d}_{12,i}^{\phantom{*}} {d}_{23,j}^{\phantom{*}} {d}_{13,k}^*$, now of rank 3, averages again to an isotropic tensor, which should be proportional to the Levi-Civita tensor $\epsilon_{ijk}$, so that $\left\langle {d}_{12,i}^{\phantom{*}} {d}_{23,j}^{\phantom{*}} {d}_{13,k}^* \right\rangle_\mathcal{O} = D_3 \, \epsilon_{ijk}$, where the multiplier $D_3$ is found by contracting with a separate Levi-Civita tensor.
This gives
\begin{align}
6\, D_3
 = D_3 \, \epsilon_{ijk}\epsilon_{ijk}
%\nonumber \\ &
& = \left\langle {d}_{12,i}^{\phantom{*}} {d}_{23,j}^{\phantom{*}} {d}_{13,k}^* \epsilon_{ijk} \right\rangle_\mathcal{O}
\nonumber \\ & = (\mathbf{d}_{12} \times \mathbf{d}_{23}) \cdot \mathbf{d}_{13}^*,
\end{align}
in terms of the orientation-invariant scalar triple product of the three dipoles. Finally, the resulting factor of $\epsilon_{ijk}$ induces the triple product $\mathcal{E}_{1,i}^{\phantom{*}} \mathcal{E}_{2,j}^{\phantom{*}} \mathcal{E}_{3,k}^* e^{i\phi}\epsilon_{ijk} = (\mathbfcal{E}_{1} \times \mathbfcal{E}_{2} ) \cdot \mathbfcal{E}_{3}^* e^{i\phi}$ for the fields, giving the final result
\begin{align}
\left\langle
|V_{12}||V_{23}||V_{13}|\cos{\phi}
\right\rangle_\mathcal{O}
& =
\frac{1}{24}
\Re\mathopen{}\left[
(\mathbfcal{E}_{1} \times \mathbfcal{E}_{2} ) \cdot \mathbfcal{E}_{3}^* e^{i\phi}
\nonumber \right. \\ & \qquad \qquad  \left.
(\mathbf{d}_{12} \times \mathbf{d}_{23}) \cdot \mathbf{d}_{13}^*
\right]\mathclose{}.
\end{align}

\subsubsection*{Ab-initio calculation of molecular dipoles}
Molecular triple products are calculated for the showcase of the  methyloxirane singly-ionised molecule (see Table~3 in SI). We treat Rydberg states within the multi-reference configuration-interaction with single excitations (MR-CIS) ansatz. The CAS(2,2) wavefunction, with the active orbitals localised on the lone pairs of the oxygen atom, was used as the reference. This calculation is performed within the ORMAS (Occupation Restricted Multiple Active Space) solver~\cite{Ivanic2003I, Ivanic2003II} of the GAMESS package~\cite{Schmidt1993, Dykstra2005} using the optimised MP2(fc) [Møller–Plesset to second order perturbation theory with frozen core] method for the geometry shown in Table~1 in SI.

We use the aug-cc-pVTZ basis set, augmented with several Kaufman-Rydberg functions (with $n=1$ through 4 and $S$, $P$, $D$ and $F$ character) at the centre of mass of the molecule to accurately support the Rydberg series of the molecule~\cite{Dunning1989, Kendall1992, Kaufmann1989}. The energies of the eigenstates of interest (discarding spin triplet states) are reported in detail in the SI.

%% file: supplementary.tex
\section*{TRICC-field configuration}
\begin{figure*}[h!]
    \centering
    \includegraphics[width=0.98\linewidth]{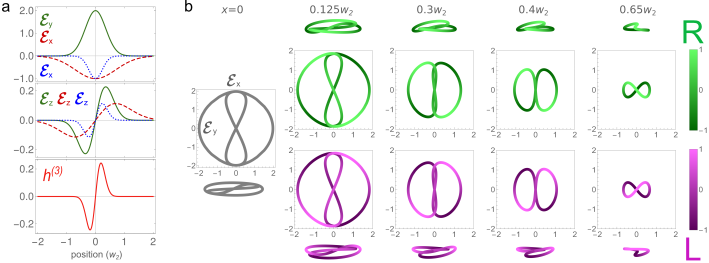}
    \caption{TRICC field.
    (a) Field components of each TRICC-field colour, both transverse (top) and longitudinal (middle), producing a nonzero chiral correlation function $h^{(3)}$ (bottom).
    The fields are colour-coded as in Figure~\ref{fig:eFIDscheme}b and normalised to the central value of the transverse component.
    (b) 3D Lissajous figures of the TRICC field, showing knotted polarisation, at different positions along the $x$ axis. The first column (gray) shows the achiral case $x=0$. For $x \neq 0$, the top two rows (green) correspond to positive values of $x$ and the bottom two rows (lilac) to negative values. The middle two rows are the projections of the Lissajous figures on the $xy$-plane, and bottom and top show an angled viewpoint. 
    The lightness of the curve (colour scales on the right) represents the value of the longitudinal component $\mathcal{E}_z(t)$.
    We show fields in a $\omega_1{:}\omega_2{:}\omega_3 = 2{:}1{:}3$ configuration with wavelengths $\lambda_1=\SI{800}{nm}$, $\lambda_2=\SI{1600}{nm}$ and $\lambda_3=\SI{533}{nm}$ and phases $\phi_1=0$, $\phi_2=\pi/2$ and $\phi_3=\pi/2$ and intensity ratio $I_1{:}I_2{:}I_3 = 4{:}1{:}1$, focused to $w_i=1.2\lambda_i$ with equal numerical aperture for all three colours.}
\label{fig:tricc_sm}
\end{figure*}

Here we present an alternative configuration of the TRICC field, shown in Figure~\ref{fig:tricc_sm} using identical conventions to Figure 2 of the main text. 
One can observe here that upon moving from the achiral middle of the beam, the Lissajous figure first acquires a complex knotted structure~\cite{Sugic2020}, which then unknots itself between $x=0.3w_2$ and $x=0.4w_2$, followed by shrinking in size as a structure, corresponding to a decrease in intensity. This behaviour appears symmetrically in both directions from $x=0$, but with opposite chirality. 
The knot here is isomorphic to the trefoil knot, which is the simplest possible chiral knot~\cite{Adams2004}.

\section*{Ab initio calculations for methyloxirane molecule}
Molecular triple products are calculated for the methyloxirane molecule. This calculation is performed within the ORMAS solver of the GAMESS package using the optimised MP2(fc), for the geometry shown in Table~\ref{tab:geometry}.

\begin{table*}[b] %% without [b] for independent use
\centering
\begin{minipage}{.58\textwidth}
  \centering
\begin{tabular}{cccc}
Atom & $x$-coordinate & $y$-coordinate & $z$-coordinate \\
\hline \hline
 C  &  -0.211761865044 &  -0.051130800700 &   0.491070227321 \\
 C  &   0.952135220773 &  -0.694687579808 &  -0.118609052200 \\
 C  &  -1.548826197176 &  -0.044575312951 &  -0.182512567928 \\
 O  &   0.813397556144 &   0.738851113898 &  -0.138422794998 \\
 H  &   1.756284129428 &  -1.062783697723 &   0.505504521687 \\
 H  &   0.836773710857 &  -1.171275968924 &  -1.084528278135 \\
 H  &  -0.214501788525 &   0.043535717665 &   1.572219470170 \\
 H  &  -2.072448641908 &   0.889377235478 &   0.018740574166 \\
 H  &  -2.163990948087 &  -0.867454009803 &   0.182557270784 \\
 H  &  -1.425212839981 &  -0.146417011606 &  -1.259309512207  
\end{tabular}
\captionof{table}{Molecular geometry of S-methyloxirane. For the R enantiomer, we spatially invert ($\mathbf{r}\mapsto -\mathbf{r}$) all atomic positions.}
\label{tab:geometry}
\end{minipage}%
\hspace{0.05\textwidth}
\begin{minipage}{.3\textwidth}
\centering
\begin{tabular}{lccc}
$n$ & Type & $E$ [eV] \\
\hline \hline 
1  & ground & 0.000 \\
3  & 3s & 7.279 \\
5  & 3p$_y$ & 7.647 \\
7  & 3p$_z$ & 7.676 \\
9  & 3p$_x$ & 7.846 \\
12 & 3d$_{z^2-x^2}$ & 8.452 \\
13 & 3d$_{z^2-y^2}$ & 8.468 \\
16 & 3d$_{xz}$ & 8.492 \\
17 & 3d$_{yz}$ & 8.517 \\
19 & 3d$_{xy}$ & 8.539
\end{tabular}
\captionof{table}{Excitation energies of neutral methyloxirane.}
\label{tab:molorb}
\end{minipage}
\end{table*}

\begin{figure*}[h]
    \vspace{1mm}
    \centering
    \includegraphics[width=0.75\linewidth]{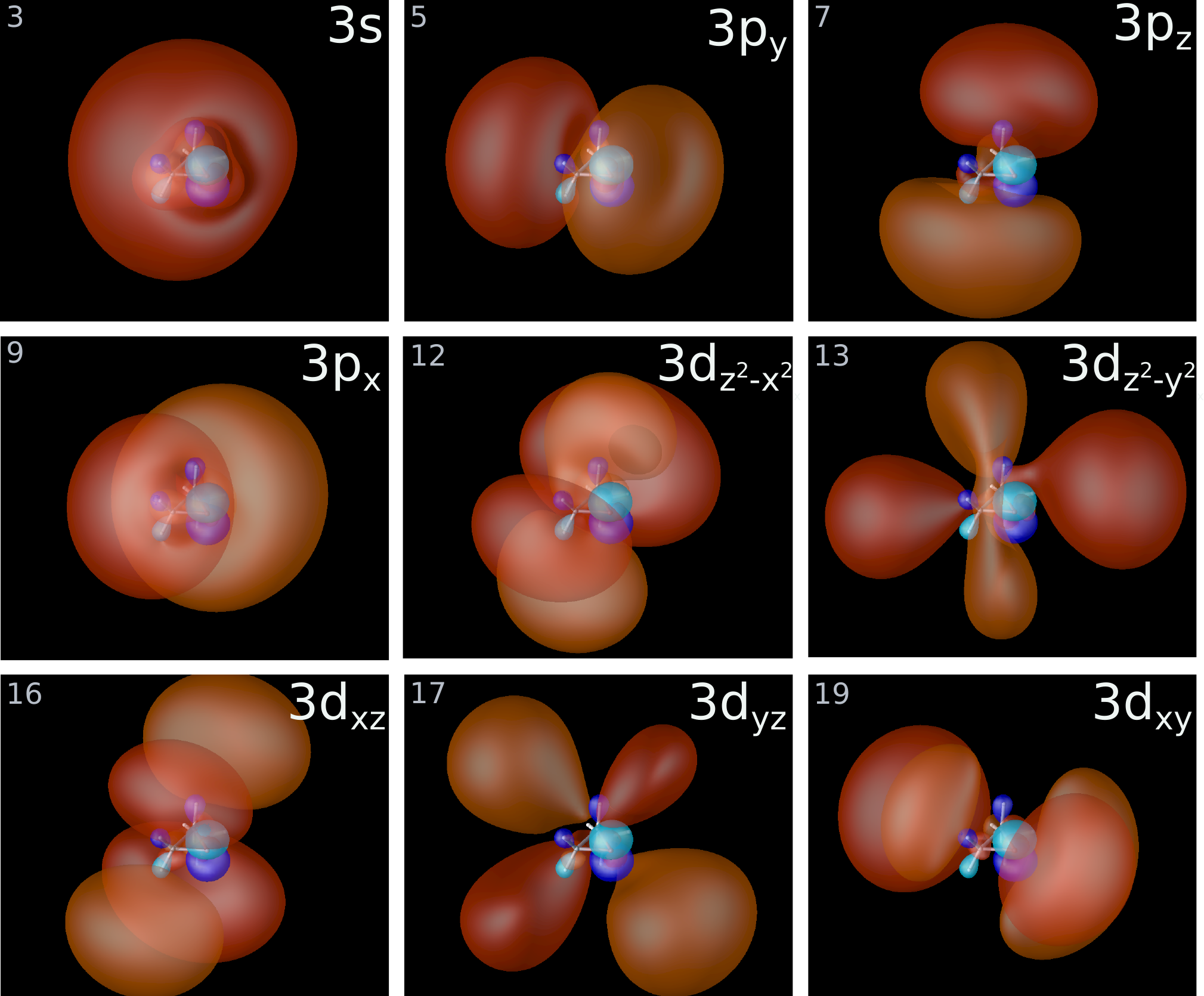}
    \caption{Molecular orbitals of Rydberg states of methyloxirane. The state indices marked on the top-left corner of each diagram correspond to the index $n$ in Table~\ref{tab:molorb}.}
    \label{fig:mo}
\end{figure*}

We use the aug-cc-pVTZ basis set, augmented with several Kaufman-Rydberg functions (with $n=1$ through 4 and $S$, $P$, $D$ and $F$ character) at the centre of mass of the molecule to accurately support the Rydberg series of the molecule. The energies of the eigenstates of interest (discarding spin triplet states) are reported in detail in Table~\ref{tab:molorb}. These are in broad agreement with previous numerical and experimental results~\cite{Barone2014}. The molecular orbitals corresponding to the excited states from this list are shown in Figure~\ref{fig:mo}.

\begin{table*}
\centering
\begin{tabular}{lccc}
$d_{i,  j}$ & $x$-component & $y$-component  & $z$-component  \\
\hline \hline \\
  $d_{1 , 1}$    &   -2.657525769894   &   -1.543353440217   &   -0.228036682970 \\
  $d_{1 , 3}$    &    0.091988977246   &    0.098002456237   &   -0.095712584970 \\
  $d_{1 , 5}$    &   -0.114785379164   &    0.004980558892   &   -0.289690684997 \\
  $d_{1 , 7}$    &    0.037337506874   &    0.212272863307   &   -0.184052734825 \\
  $d_{1 , 9}$    &   -0.164384462658   &    0.121154631329   &   -0.053858352076 \\
  $d_{1, 12}$    &    0.070286207770   &    0.069626845409   &    0.037556561082 \\
  $d_{1, 13}$    &   -0.074199408815   &   -0.013384363110   &    0.010511482901 \\
  $d_{1, 16}$    &    0.027297040930   &    0.092264353832   &    0.103006395326 \\
  $d_{1, 17}$    &    0.057685396468   &    0.042479894136   &   -0.040279222614 \\
  $d_{1, 19}$    &    0.095843990557   &    0.054546484613   &    0.070448835820 \\
  $d_{3,  3}$    &   -1.448749800607   &   -2.470471844118   &    1.249356231590 \\
  $d_{3,  5}$    &   -1.114511350645   &   -3.947903746065   &    0.334427377246 \\
  $d_{3,  7}$    &   -0.558991208987   &    0.809582255126   &    4.517508937859 \\
  $d_{3,  9}$    &    4.114951108384   &   -1.182732295471   &    0.590975182563 \\
  $d_{3, 12}$    &    0.168949478830   &   -0.350672741671   &    0.037572073151 \\
  $d_{3, 13}$    &    0.446935313569   &    0.483329506287   &   -0.373947285238 \\
  $d_{3, 16}$    &    0.363094365047   &    0.003900424116   &    0.164417081654 \\
  $d_{3, 17}$    &   -0.038559865609   &    0.021972944780   &   -0.104292850068 \\
  $d_{3, 19}$    &   -0.092595456686   &    0.055480939162   &   -0.202647782194 \\
  $d_{5 , 5}$    &   -2.180684792012   &   -2.968182981068   &    0.101798958996 \\
  $d_{5 , 7}$    &   0.254188557147    &   0.480999249790    &  -0.253778457339 \\
  $d_{5 , 9}$    &   -0.104381185028   &    0.986303830685   &   -0.007190547362 \\
  $d_{5, 12}$    &   -0.711881012677   &   -0.084504239273   &    0.625397520615 \\
  $d_{5, 13}$    &    1.533777013099   &   -3.101539229162   &   -1.246498312830 \\
  $d_{5, 16}$    &   -0.551530989191   &   -0.240930033441   &   -0.536682494495 \\
  $d_{5, 17}$    &   -0.378927665922   &    1.268416459796   &   -3.314499363107 \\
  $d_{5, 19}$    &   -2.848550143508   &   -2.313210974351   &   -0.012329313169 \\
  $d_{7 , 7}$    &   -2.328870334263   &   -2.285466458866   &   -1.270255263971 \\
  $d_{7,  9}$    &   -0.509338021447   &   -0.098476829478   &   -1.107049326475 \\
  $d_{7, 12}$    &   -1.237545945677   &   -0.541292624163   &   -2.233025481185 \\
  $d_{7, 13}$    &    0.137385765355   &    1.504056651851   &   -2.133313540053 \\
  $d_{7, 16}$    &    3.605808682847   &   -0.692177595294   &   -1.260955079547 \\
  $d_{7, 17}$    &    0.208138130879   &    3.323895563713   &    0.458193512005 \\
  $d_{7, 19}$    &   -0.354540262592   &    0.022282934444   &   -1.301367228212 \\
  $d_{9 , 9}$    &   -3.874288814724   &   -1.550375724064   &    0.058365661961 \\
  $d_{9, 12}$    &    4.935063470291   &   -0.657198772570   &   -1.076021479885 \\
  $d_{9, 13}$    &   -0.535928496260   &   -2.796788649119   &   -0.201588961480 \\
  $d_{9, 16}$    &    1.045256445603   &    0.880354995134   &    3.983950035917 \\
  $d_{9, 17}$    &    1.150385445327   &    1.263419868559   &   -0.768887170659 \\
  $d_{9, 19}$    &   -1.465435303259   &    3.108773778035   &   -1.192956486852 \\
 $d_{12, 12}$    &   -5.089175393465   &   -2.443470348450   &   -0.298934325837 \\
 $d_{12, 13}$    &    1.623607969100   &    0.301040670738   &   -0.800995671608 \\
 $d_{12, 16}$    &   -0.758543852920   &   -0.045759323038   &   -0.562841534292 \\
 $d_{12, 17}$    &   -0.810318121435   &    0.327913759159   &   -0.144181702084 \\
 $d_{12, 19}$    &    0.324395698475   &   -1.135479304243   &   -0.053931325799 \\
 $d_{13, 13}$    &   -4.182746411120   &   -1.014375241188   &    0.905380972497 \\
 $d_{13, 16}$    &    0.156653112160   &    0.175851142115   &    0.927768518583 \\
 $d_{13, 17}$    &    0.620771731324   &   -1.104818811914   &    0.928724046325 \\ 
 $d_{13, 19}$    &    1.053932021966   &    0.598515893605   &   -0.356018369896 \\
 $d_{16, 16}$    &   -4.295792962226   &   -2.312435677284   &   -0.630725463029 \\
 $d_{16, 17}$    &   -0.543105275103   &   -0.794254672959   &   -0.373800186865 \\
 $d_{16, 19}$    &    0.158664049166   &   -0.502310230162   &    0.770209037724 \\
 $d_{17, 17}$    &   -3.688302409687   &   -1.917649666414   &   -1.585518036329 \\
 $d_{17, 19}$    &   -0.569701898997   &   -0.370546016532   &   -0.502881920693 \\
 $d_{19, 19}$    &   -4.497206718969   &   -2.156411622373   &   -0.418743285228 
\end{tabular}
\caption{Calculated values of transition dipole moment components, in atomic units.}
\label{tab:num_dips}
\end{table*}

The calculated transition dipoles $d_{i,j}$ for all states under consideration are given in Table~\ref{tab:num_dips}. Excitation to all of these states is dipole-allowed as expected for a C$_1$ molecule.
For added clarity, we also present the vector triple products of transition dipoles between ($3s,3p,3d$)-states. In Table~\ref{tab:triple_dips} we show the major values of these triple-dipole products.

\begin{table*}[h!]
\centering
    \begin{tabular}{c|ccccc}
          &    12     &    13    &    16    &     17      &     19 \\
  State   & $3d_{z^2-x^2}$ & $3d_{z^2-y^2}$ &  $3d_{xz}$   &    $3d_{yz}$   &    $3d_{xy}$ \\
\hline \\[-0.5em]
  5 \quad  $3p_y$  & -0.675  &    -1.318   &    0.482  &   -0.269   &    1.627 \\
  7 \quad  $3p_z$  &  2.555  &    -3.729   &    0.409  &    0.802   &  -0.078 \\
  9 \quad  $3p_x$  & -2.177  &     5.634   &   -1.162  &  -0.606    &   -1.977 
\end{tabular}
\caption{Values of molecular triple-dipole products. The lowest $3s$-state is common for each product.}
\label{tab:triple_dips}
\end{table*}

$\ $

$\ $

$\ $

\section*{Benchmarking of TRICC dynamics in methyloxirane}
\begin{figure}[ht!]
    \vspace{3mm}
    \centering
    \includegraphics[width=0.45\linewidth]{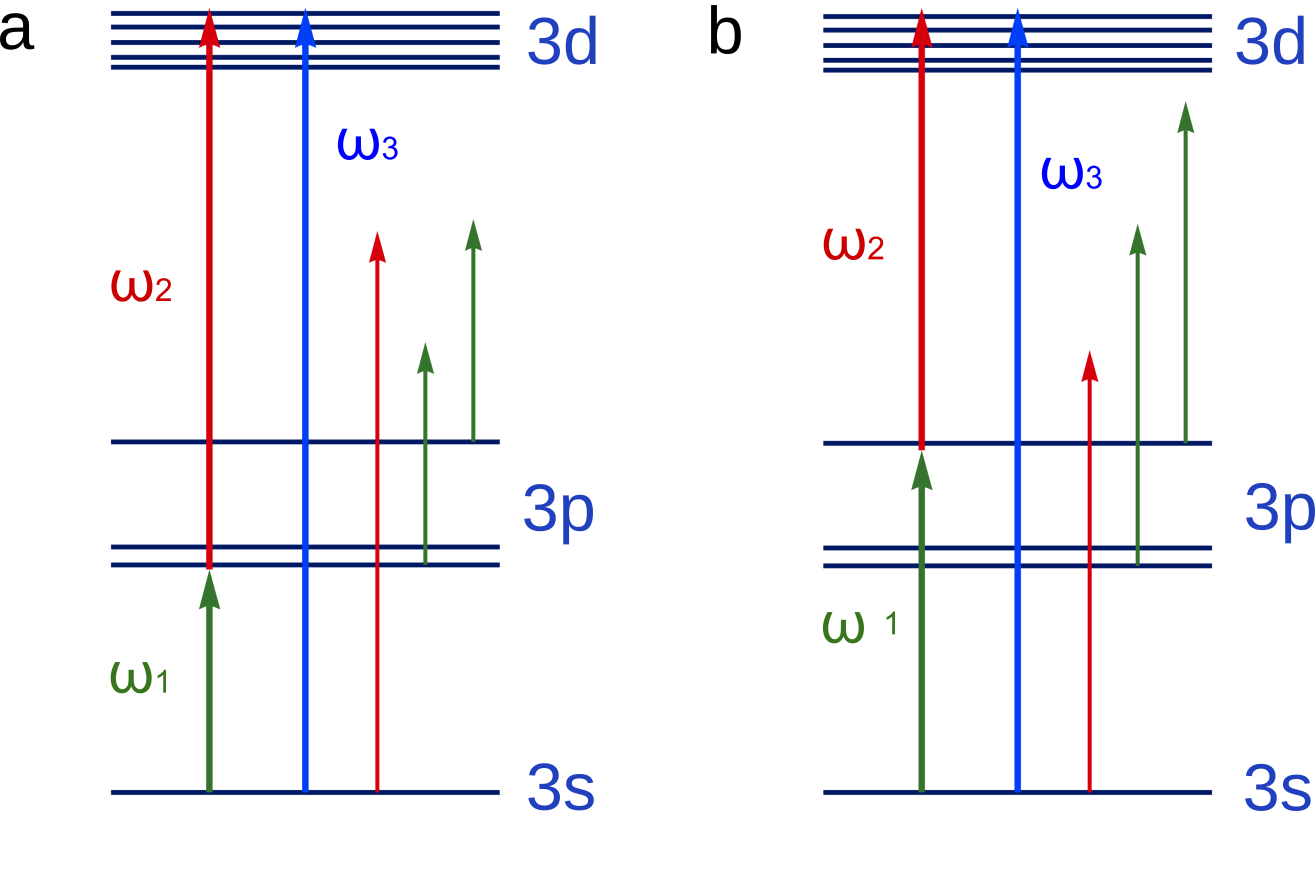}
    \caption{Schemes of methyloxirane driven by TRICC field with wavelengths (a) $\lambda_1=$ \SI{3438}{nm}, $\lambda_2=$ \SI{1365}{nm}, and $\lambda_3=$ \SI{977}{nm}, or (b) $\lambda_1=$ \SI{2231}{nm}, $\lambda_2=$ \SI{1726}{nm}, and $\lambda_3=$ \SI{973}{nm} close to resonances for transitions between $3s$, (a) $3p_y$ and (b) $3p_x$, and $3d_{xy}$ states with energies from Table~\ref{tab:molorb}.}
    \label{fig:schemes}
\end{figure}

To understand how visible the eFID effect can be in the `real' system, we consider two schemes shown in Figure~\ref{fig:schemes}, where all the states with their energies and transition dipoles are taken from the \emph{ab-initio} calculations. 

The two schemes differ from each other by the choice of which $3p$ state is closest to the resonance (either the lowest, $3p_y$, or the highest, $3p_x$), but with the same $3d$ state, $3d_{xy}$, being closest to the resonance. These two schemes realise situations when values of both (i) the triple molecular-dipole product and (ii) the triple TRICC-field product are as large as possible, but still ensure that the molecule has population left in the FID-active state ($|1 \rangle$ or here $3s$) by the end of the TRICC pulse.

\begin{figure}[t] %% [ht!] for independent use
    \vspace{3mm}
    \centering
    \includegraphics[width=0.98\linewidth]{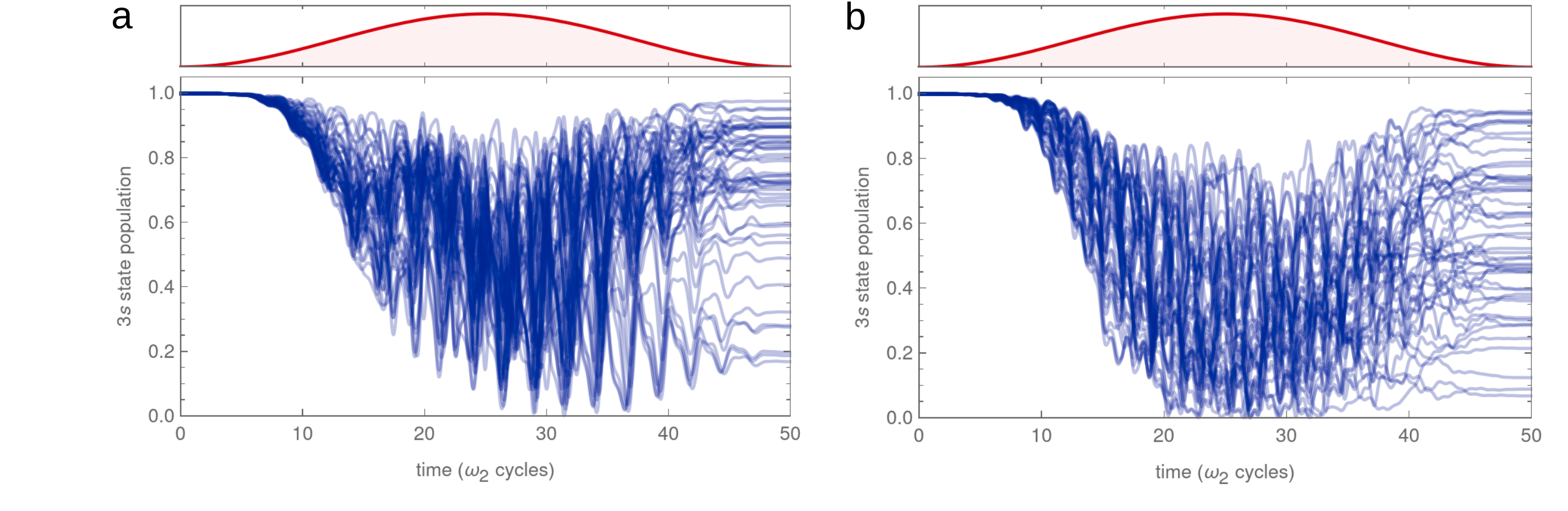}
    \caption{Population of the FID-active $3s$ state of methyloxirane for a random sample of orientations of the molecule during the TRICC pulse. The wavelengths of the TRICC field components correspond to the schemes in Figure~\ref{fig:schemes} (a) and (b), respectively. The intensities of the TRICC field components are (a) $I_1= \SI{2e10}{W/cm^2}$, $I_2= \SI{3.5e11}{W/cm^2}$ and $I_3= \SI{2e11}{W/cm^2}$, and (b) $I_1= \SI{1.5e10}{W/cm^2}$, $I_2= \SI{1e11}{W/cm^2}$ and $I_3= \SI{1.2e11}{W/cm^2}$; the corresponding phases are $\phi_1=\pi/3$, $\phi_2=-\pi/3$, and $\phi_3=\pi$. The pulse duration (intensity FWHM) is $25$ cycles of the $\omega_2$ field, and the focal waist of the UV beam is (a) $w_{UV}=7\lambda_{UV}$ and (b) $w_{UV}=6\lambda_{UV}$. The top panel shows the $\sin^2$ envelope of the TRICC field.}
    \label{fig:benchmarking_population}
\end{figure}

Figure~\ref{fig:benchmarking_population} shows the numerically-calculated population of the FID-active $3s$ state during the TRICC-field pulse for different orientations of the molecule, for both schemes shown in Figure~\ref{fig:schemes}. Here one can see that the majority of orientations presents relatively high population of the $3s$-state after the pulse, while a minority end up with down to (a) 20\% and (b) 5\% population, which still secures the final effect. 
These simulations are obtained by direct solution of the TDSE (see Methods) in the full system of states listed in Table~\ref{tab:molorb} with the dipole moments listed in Table~\ref{tab:num_dips}, using the standard numerical ODE integration functions of the \textsc{Mathematica} software package.

We benchmark the phase behaviour of the FID-active $3s$ state for a range of intensities up to those used in Figure~\ref{fig:schemes}. We compare the phase found analogously to Eq.~\eqref{off_res_en_shift} of the main text, but taking into account all of the states shown in Figure~\ref{fig:schemes}, as
\begin{equation}
\langle \delta E \rangle_{\mathcal{O}} = \sum_{\mathrm{pd}} \frac{\Re  \mathopen{} \left\{ \left( \mathbf{d}^*_{\mathrm{sd}} \cdot \left[ \mathbf{d}_{\mathrm{sp}} \times \mathbf{d}_{\mathrm{pd}} \right] \right) \left( \mathbfcal{E}^*_3 \cdot \left[\mathbfcal{E}_1 \times \mathbfcal{E}_2 \right] e^{i\phi} \right)\mathclose{} \right\}}{24 \omega_{\mathrm{sp}} \omega_{\mathrm{sd}}}  
+\sum_{\mathrm{p}} \frac{|\mathbf{d}_{\mathrm{sp}}|^2 |\mathbfcal{E}_1|^2}{12\omega_{\mathrm{sp}}} + \sum_{\mathrm{d}} \frac{ |\mathbf{d}_{\mathrm{sd}}|^2 |\mathbfcal{E}_3|^2}{12 \omega_{\mathrm{sd}}}  \, ,
\label{off-res_en_shift_aver_full}
\end{equation}
and using the \emph{ab initio} molecular dipoles from Table~\ref{tab:num_dips}, with the numerical solution of the system of differential equations, further combined with orientation averaging. The orientation averaging for the numerical TDSE solution is done numerically, using Fibonacci numerical integration on a sphere~\cite{Hannay2004} (with improved implementation as per Ref.~\cite{FibonacciSE}) for the orientation rotation axis, and cyclic rectangle-rule integration for the orientation rotation angle.

\begin{figure}[ht!]
    \vspace{3mm}
    \centering
    \includegraphics[width=0.98\linewidth]{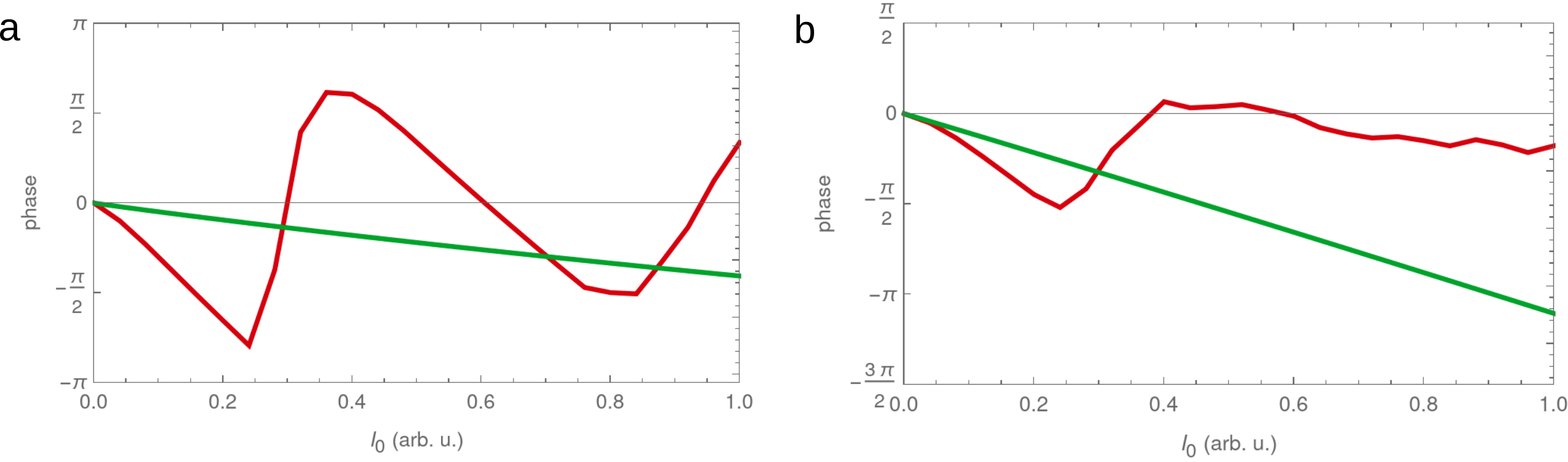}
    \caption{The accumulated phase of the FID-active $3s$ state due to the TRICC dynamics,  calculated fully analytically~(\ref{off-res_en_shift_aver_full}) (green) and numerically (red), as a function of global intensity (i.e.\ each intensity used for Figure~\ref{fig:benchmarking_population} is multiplied by $I_0$).}
    \label{fig:benchmarking_phase}
\end{figure}

Figure~\ref{fig:benchmarking_phase} shows that there is a qualitative agreement between the numerical and analytical~(\ref{off-res_en_shift_aver_full}) solutions averaged over molecular orientations for low intensities. For higher intensities the numerical phase starts oscillating around the analytical one, presenting partially resonant features. However, the values of the phase for both calculations are numbers of the same order and should result in the same strong effect.
Moreover, to the extent that the calculations disagree, the analytical result used in the main text is an under-estimation of the numerical-TDSE result, which indicates that the results reported in the main text should, in real-world experiments, be achievable using lower driver intensities.

\section*{Additional results}
\begin{figure}[ht!]
    \vspace{3mm}
    \centering
    \includegraphics[width=0.45\linewidth]{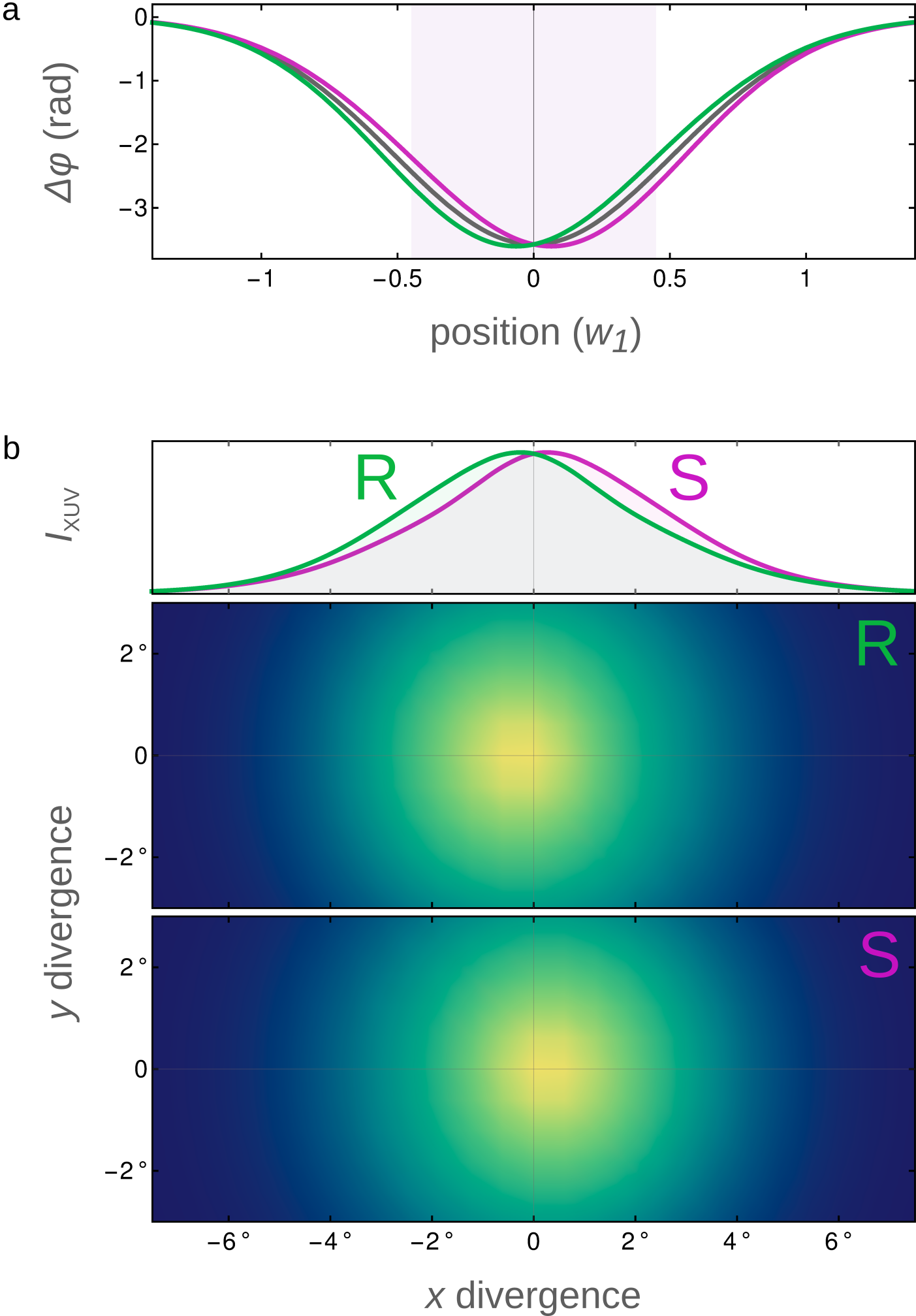}
    \caption{eFID by methyloxirane. 
    (a) Phase $\Delta \varphi$ accumulated in the FID-active Rydberg $3s$ state for R (green) and  S (lilac) enantiomers; gray shows an achiral phase. 
    (b) FID beam divergence for each enantiomer, with a lineout of both on the top panel. 
    The beams are deflected by about \SI{0.5}{degrees}. We use TRICC-field parameters corresponding to Figures~\ref{fig:schemes}b and~\ref{fig:benchmarking_population}b, with the TRICC field components focused down to equal waists $w_1=w_2=w_3=1.2\lambda_1$.
    }
    \label{fig:real_res_2}
\end{figure}

Here we present the result for the enantiosensitive steering of FID by methyloxirane, which corresponds to the second scheme considered above (see Figure~\ref{fig:schemes}b), with the first scheme shown in the main text. We use TRICC-field parameters from Figure~\ref{fig:benchmarking_population}b. Figure~\ref{fig:real_res_2}a presents the phase of the FID-active $3s$ state accumulated during the TRICC pulse for different enantiomers of methyloxirane. The resulting deflection of the FID UV beam is presented in Figure~\ref{fig:real_res_2}b for both enantiomers.
One can see that each enantiomer sends the beam at half a degree from the initial direction, which is easily observable experimentally.

\newpage

\section*{Resonant case}
Here we derive an analytical solution for the TDSE (see Methods) for the resonant case where there is zero detuning between the driving lasers and the corresponding molecular electronic transitions.
We start from reducing the number of equations in the system~(see Methods) by eliminating the complex amplitude of the state $|3 \rangle$, which brings us to a system
\begin{equation}
\begin{split}
i \ddot{c}_1 - \omega_{13} \dot{c}_1 & + i |V_{13}|^2 c_1 = - V_{12} e^{-i\omega_{12} t}\dot{c}_2 
 - i \left(  V_{13} V^*_{23} + \omega_{23} V_{12}\right) e^{-i\omega_{12} t} c_2  \, ,\\
i \ddot{c}_2 - \omega_{13} \dot{c}_2 & + i |V_{13}|^2 c_2 = - V^*_{12} e^{i\omega_{12} t}\dot{c}_1 
 - i \left( V^*_{13} V_{23} + \omega_{13} V^*_{12}\right) e^{i\omega_{12} t} c_1 \, .
\end{split}
\label{general_system_reduced}
\end{equation}
For nonzero detunings, this system of differential equations does not have an analytical solution.
However, for the resonant case, this system simplifies to
\begin{equation}
\begin{split}
i \ddot{c}_1 + i |V_{13}|^2 & c_1 = - V_{12} \dot{c}_2 - i  V_{13} V^*_{23} c_2  \, ,\\
i \ddot{c}_2 + i |V_{13}|^2 & c_2 = - V^*_{12} \dot{c}_1 - i V^*_{13} V_{23} c_1 \, ,
\end{split}
\label{res_system}
\end{equation}
which can be solved within the ansatz $c_{1,2}=\tilde{c}_{1,2} e^{-\lambda t}$. The corresponding characteristic equation is 
\begin{equation*}
\begin{split}
\lambda^4+\lambda^2(|V_{12}|^2+|V_{13}|^2+&|V_{23}|^2) 
-i\lambda(V_{12}  V_{13}^* V_{23} + V^*_{12} V_{13} V^*_{23}) = 0 \, ,
\end{split}
\end{equation*}
with solutions
\begin{equation}
\lambda = \frac{2^{1/3}a}{\gamma} 
\begin{Bmatrix}
e^{i\pi} \\
e^{i\pi/3} \\
e^{-i\pi/3} \\
0
\end{Bmatrix}
-\frac{\gamma}{2^{1/3}}
\begin{Bmatrix}
e^{-i\pi} \\
e^{-i\pi/3} \\
e^{i\pi/3} \\
0
\end{Bmatrix} \, ,
\label{res_lamdas}
\end{equation}
where 
\begin{equation}
\begin{split}
&a = \frac{1}{3}(|V_{12}|^2+|V_{13}|^2+|V_{23}|^2) \, , \\
&\gamma = \left(ib +\sqrt{4a^3-b^2} \right)^{1/3} \, , \\
&b = V_{12} V_{13}^* V_{23} + V^*_{12} V_{13} V^*_{23} = 2 |V_{12}| |V_{13}| |V_{23}| \cos{\phi} \, , \\
&\phi = \phi_1 + \phi_2 - \phi_3 \, .
\end{split}
\label{res_notations}
\end{equation}
The trivial solution here addresses the stationary case, which is realised when $V_{12} V_{13}^* V_{23} = - V^*_{12} V_{13} V^*_{23}$ corresponding to $\phi=\pi/2$.

If one of the interaction matrix elements~(see Methods) is much smaller than other two, then the solutions $\lambda$ can be written in the form:
\begin{equation}
\lambda = 2i \left( \sqrt{a} \sin{\phi_0} -\frac{b}{6a} \cos{\phi_0}\right) \, ,
\label{res_lamdas_simp}
\end{equation}
where $\phi_0$ is a constant phase for the different values of $\lambda_{1,2,3}$ in~(\ref{res_lamdas}). The complex amplitude of the state $|1 \rangle$ in this case takes the form $c_1(t)=e^{i \delta E t}$, where
\begin{equation}
\delta E =  \frac{2 |V_{12}||V_{13}||V_{23}| }{|V_{12}|^2+|V_{13}|^2+|V_{23}|^2} \cos{\phi} \cos{\phi_0} -\frac{2}{\sqrt{3}}(|V_{12}|^2+|V_{13}|^2+|V_{23}|^2)^{1/2} \sin{\phi_0} \, .
\label{res_en_shift}
\end{equation}
One can notice that the fact that $\lambda$ in~(\ref{res_lamdas_simp}) is imaginary leads to a pure energy shift of the excited state $|1 \rangle$. Moreover, this energy shift (equivalently, phase shift) includes a linear dependence on the triple product of interaction matrix elements, which can be controlled through the relative phase $\phi$ between the TRICC field components.

From the various solutions we found above in~(\ref{res_lamdas}) associated with different values of $\phi_0$, we are interested in the solution corresponding to the situation where the initial population before the TRICC pulse is in the FID-active state,~$|1 \rangle$. In our case it is reasonable to assume that the longest wavelength field, $\mathcal{E}_2$, starts first, in which case the solution of interest corresponds to $\phi_0=\pi$. In this case, the energy shift simplifies to
\begin{equation}
\delta E = - \frac{2 |V_{12}||V_{13}||V_{23}| }{|V_{12}|^2+|V_{13}|^2+|V_{23}|^2} \cos{\phi} \, ,
\label{res_en_shift_phi0_pi}
\end{equation}
and, in the limit of $|V_{23}| \gg |V_{12}|,|V_{13}|$, to
\begin{equation}
\delta E = - \frac{2 |V_{12}||V_{13}|}{|V_{23}|} \cos{\phi} \, .
\label{res_en_shift_phi0_pi}
\end{equation}
However, in this case the orientation averaging can only be approached numerically, due to the presence of $|V_{23}|$ in the denominator. (Moreover, the approximation $|V_{23}| \gg |V_{12}|,|V_{13}|$ cannot hold uniformly for all molecular orientations, since $|V_{23}|$ depends on an inner product with $\mathbfcal{E}_2$.) This raises the difficulty of analysis for this case as well as the numerical cost of computation.